\documentclass{article}
\usepackage{graphicx}
\title{Quantum Portfolios of Observables and the Risk
Neutral Valuation Model}
\author{Fredrick Michael* \\ {\it {written Nov 13 2003}}}

\begin{document}
\maketitle
\begin{abstract}

Quantum Portfolios of quantum algorithms encoded on qbits have recently been reported. In this paper a discussion of the continuous variables
version of quantum portfolios is presented. A risk neutral valuation model
for options dependent on the measured values of the observables, analogous to the traditional Black-Scholes valuation model, is obtained from
the underlying stochastic equations. The quantum algorithms are here
encoded on simple harmonic oscillator (SHO) states, and a Fokker-Planck
equation for the Glauber P-representation is obtained as a starting point
for the analysis. A discussion of the observation of the polarization of a
portfolio of qbits is also obtained and the resultant Fokker-Planck equation is used to obtain the risk neutral valuation of the qbit polarization
portfolio.
\end{abstract}

\section{Introduction}
Recently, quantum portfolios of quantum algorithms encoded on two-level
systems (qbits) have been reported \cite{a}. It has been shown that maximizing the
efficiency of quantum calculations can be accomplished via the formation of
portfolios of the algorithms and by minimizing both the running time and its
uncertainty. Furthermore, these portfolios have been shown to outperform single
algorithms when applied to certain types of algorithms with variable running
time such as the "Las Vegas" algorithms and NP-complete problems such as
combinatorial search algorithms.
In this paper a discussion of quantum portfolios of algorithms encoded on
qbits and their continuous variables (CV) version (here,harmonic oscillators)
is presented. Continuous variables quantum computation performed with harmonic oscillators is discussed in \cite{b,c,d}. A quantum portfolio of algorithms
is formed and a risk neutral model, analogous to the traditional Black-Scholes
and Merton \cite{e,f,g} options valuation model, is obtained from the underlying
stochastic equations. The quantum algorithms are here encoded on simple harmonic oscillator (SHO) states, and a Fokker-Planck equation for the Glauber
P-representation is obtained as a starting point for the analysis. A portfolio
of qbits is also formed and the resultant Fokker-Planck equation of the qbit
polarization is used to obtain the risk neutral valuation of the portfolio and
measurement option. The results should prove useful in quantum computation
and decoherence.

\section{Preliminaries}
A simplified model for quantum computation is proposed wherein the algorithms
are encoded on simple harmonic oscillator basis states and are in the presence
of a thermal bath, the temperature decoherence effects needing to be taken
into account as is usually the case in real-world applications. The quantum
computation utilizing harmonic oscillator basis sets representations has been
discussed elsewhere \cite{b,c}. Representations of computation operators or effective
operators can be made and the basis sets of these operators can be expressed as
harmonic oscillator basis sets. For this case, we will also consider damping, this could be from other noise sources, external or system specific. The harmonic  oscillator can be described by a master equation such as 
\begin{equation}
\includegraphics[width=110mm]{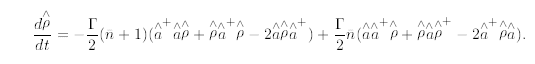} \label{eqn1}
\end{equation}
Here, $\hat a$ and $\hat a^\dagger$ are destruction and creation operators as usual, and $\hat \rho$ is the density matrix operator. The mean thermal excitation due to the thermal bath is $n=(e^{\frac{\hbar \omega} { {k_B} T         } } -1)^{-1}$, and $\Gamma$ is the damping rate \cite{h}. To obtain a Fokker-Planck equation for the SHO, we first suppose that $\hat \rho$ has a Glauber P-respresentation \cite{h}
\begin{equation}
\includegraphics[width=70mm]{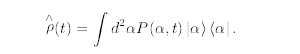} \label{eqn2}
\end{equation}
Substituting Eq.(\ref{eqn2}) into Eq.(\ref{eqn1}) we obtain the Fokker-Planck equation with mixed derivatives
\begin{equation}
\includegraphics[width=110mm]{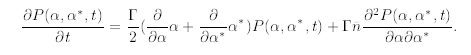} \label{eqn3}
\end{equation}
The Fokker-Planck equation is put into a regular form if we change variables using quadratures $\alpha=x+iy$, and Eq.(\ref{eqn3}) becomes
\begin{equation}
\includegraphics[width=110mm]{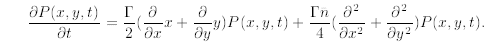} \label{eqn4}
\end{equation}

\section{Risk Neutral Valuation}
The Fokker-Planck equation for the SHO implies that there is an underlying 
stochastic differential equation(s) of the Ito form \cite{h}
\begin{equation}
\includegraphics[width=59mm]{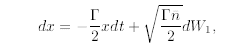} \label{eqn5}
\end{equation}
\begin{equation}
\includegraphics[width=59mm]{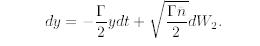} \label{eqn6}
\end{equation}

Here, the Wiener processes $dW$ (noise) are taken to be a Gaussian white noise, and are delta correlated. These stochastic processes can now be included in a portfolio analysis. To construct a portfolio, we first look at the case where the algorithm is encoded on two harmonic oscillators. It can be seen that a generalization to the multi-asset case proceeds straightforwardly. We write the Legendre transform for the N-asset portfolio $\Pi$
\begin{equation}
\includegraphics[width=100mm]{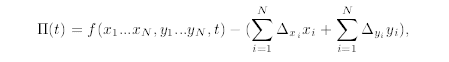} \label{eqn7}
\end{equation}
and here, with $N=2$ , the function $ f(x_1 ,x_2 , y_1 , y_2 , t)$ is the analogue to the (call) option in finance. In our case, it can represent the probability distribution for measurement of the observables of the quantum algorithm(s). These observables are Legendre transformed such that the state function, the portfolio $\Pi$ , evolves at the known or {\it{wanted}}  rate $r$ such that $d\Pi=r\Pi dt$.
The N algorithms evolve independently, regardless of 
initial phase, and will have differing instantaneous values in their stochastic
sample paths. They all have in common the same value for their moments 
(means and variances), as from the point of view of their statistics they have 
the same drift and diffusion coefficients and are equivalent on the average. This 
is different from the financial portfolio case where the portfolio of shares of a stock(s)  assets have the same instantaneous 
value for all the shares (say, of a particular stock) and would resemble a portfolio of different traded stocks of one share per stock that are different in prices at any instantaneous period in time yet that equivalently have the same values for their 1st and second moments of means and variances.

Next we expand $df$ 
in a Taylor series to first order in time, and to 2nd order in the variables, since $dW^2 =dt$
\begin{equation}
\includegraphics[width=120mm]{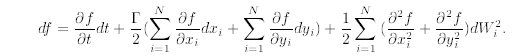} \label{eqn8}
\end{equation}
Setting $N=2$ for the two asset case, we take the differentials for $d\Pi$. We have

\begin{equation}
\includegraphics[width=70mm]{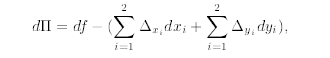} \label{eqn9}
\end{equation}
and we substitute the expressions for $df$ and for $dx_i 's$ and $dy_i 's$ . Thus we are able to make the choice for the $\Delta_{x_i ,y_i} $ multipliers directly \cite{e,f} as the $\frac{\partial f}{\partial x_i}$ terms.
 
This Immediately eliminates the drift terms in the original process, in favor of  the known rate $r$ drift terms. This is a general feature in the sense that much more complex (nonlinear) process drifts can be eliminated in favor of the known rate $r$ linear drifts \cite{g}. Upon making the substitutions, we obtain the backwards Fokker-Planck equation that corresponds to the Black-Scholes options pricing model. This PDE for the two-asset (and similarly for N$>$2 algorithms) portfolio is

\begin{equation}
\includegraphics[width=80mm]{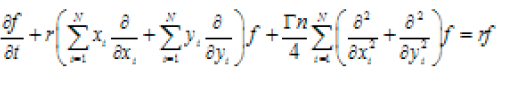} \label{eqn10} \label{10}
\end{equation}
The two-point version of this equation can be solved with the following choice for the function 
 \begin{equation}
\includegraphics[width=100mm]{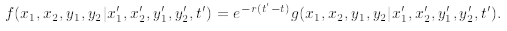} \label{eqn11} \label{11}
\end{equation}
With this choice, the PDE is transformed to a standard form backwards Fokker-Planck PDE. 
This backwards PDE is solved by $g$. Here the two-point function $g$ defined in Eq.(\ref{11}) also solves the forward Fokker-Planck equation
 \begin{equation}
\includegraphics[width=90mm]{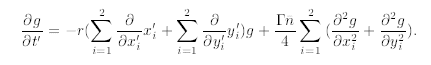} \label{eqn13} \label{13}
\end{equation}
Thus the underlying stochastic processes are of the form

 \begin{equation}
\includegraphics[width=41mm]{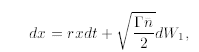} \label{eqn14} \label{14}
\end{equation} 
 \begin{equation}
\includegraphics[width=41mm]{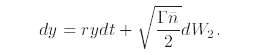} \label{eqn15} \label{15}
\end{equation} 

The solution of the one-point function $g$ is a Gaussian. We obtain the solution by transforming the stochastic differential equations to a drift-free $dx => d z_1 ,dy => d z_2$ form. The resultant one-point function $f(z_1,z_2 ,t)$ is then

\begin{equation}
\includegraphics[width=60mm]{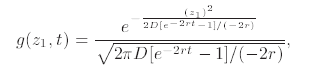} \label{eqn16} \label{16}
\end{equation} 
and similarly for the other variable $z_2$.  Here the diffusion constant is related to the damping rate as $D=\frac{\Gamma n}{2}$. The observables statistics 2-point function  is formally solved by

\begin{equation}
\includegraphics[width=120mm]{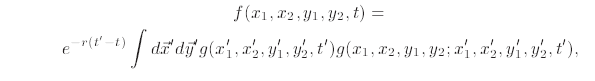} \label{eqn17} 
\end{equation} 
and we can make the boundary value choices for the observables at the later time $t'$
 
\begin{equation}
\includegraphics[width=100mm]{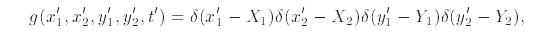} \label{eqn19a} 
\end{equation} 
where these boundary values correspond to the terminal values of the evolution of the algorithms. These boundary values allow us to immediately integrate and obtain the solution for $t'=T$
\begin{equation}
\includegraphics[width=100mm]{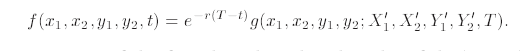} \label{eqn19} 
\end{equation} 
The interpretation of this formal result for the delta function boundary values is that the value of the distribution of the observables of the quantum computation algorithm is evaluated at the later time $t'=T$ and then evolved backwards to the present time $t$ similarly to the evaluation of the price of a financial instrument such as the call or put option in portfolio theory. Other terminal boundary values can be made, such as step functions in direct analogy to insurance instruments. Closed form solutions are also possible, as the step function limits the integration and can be solved by several methods, and will not be discussed further here...note that due to risk management strategies in portfolio theory having a direct correspondence to observation strategies for portfolios of quantum computation algorithms, alternative boundary values are likely to become important for different applications. Also, as there are generalizations of the portfolio theory to 'beyond' the risk neutral valuation model, the simple delta function boundary values and formal solution can be considered as a simplest observation strategy 
.
An example of the implementation of the strategy of observations of the quantum computations is the analog of the delta hedging strategy. The deltas $\frac{\partial f}{\partial x_i}$ and $\frac{\partial f}{\partial y_i}$ should be kept as close to zero as possible to ensure the evolution of the portfolio as $d\Pi=r \Pi dt $. This can be done by observing algorithms (selling) or adding algorithms (buying) to the portfolio at each computation time step.

\section{Polarization of qbits and Fokker-Planck equations }

In a paper that describes realistic models of interferometric continuous measurement of qbit readout systems relevant to quantum computing \cite{i}, the author
 was able to describe the polarization of the qbit in terms of a Fokker-Planck 
PDE that accounted for the randomness inherent in the measurement process. 
 The result due to the interferometric measurement Markov process was given 
by (a similar analysis can be performed for the qbits in the presence of thermal 
noise) 
\begin{equation}
\includegraphics[width=90mm]{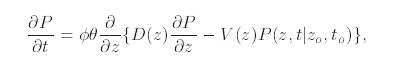} \label{eqn20} 
\end{equation} 
here $D(z)={(1- z^2 )^2} /2$ is the diffusion coefficient and $V(z)=2z(1- z^2)$ is the drift velocity. The microscopic qbit polarization yields the time dependent polarization $Z(t) (-1 < z(t)<1)$ since here 

\begin{equation}
\includegraphics[width=70mm]{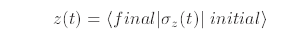} \label{eqn21} 
\end{equation} 
is the 
 polarization 
from Eq.(\ref{eqn21}) .  
The other terms in Eq.(\ref{eqn20}) are $\phi$, the photon flux in photons per second, and $\theta(order(10^-6))$ , the phase shift due to perturbation of the qbit by a photon. The underlying Ito stochastic differential equation can be readily obtained and is 
\begin{equation}
\includegraphics[width=90mm]{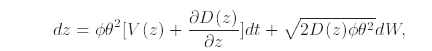} \label{eqn2351} 
\end{equation} 
and can be seen to contain complicated drift terms as well as the diffusion coefficients. This however will be the replacement by a linear drift term mentioned in the previous derivation. The portfolio analysis proceeds as before and we obtain for a two-qbit measurement distribution the following Black-Scholes PDE
\begin{equation}
\includegraphics[width=90mm]{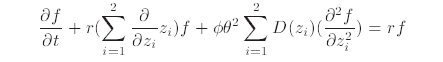} \label{24} 
\end{equation} 
Defining the function $f$ as $f=e^{-r(T-t)} g$ and substituting, we obtain the PDE 

\begin{equation}
\includegraphics[width=90mm]{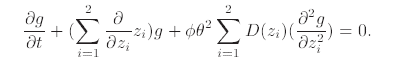} \label{25} 
\end{equation}
This equation implies that the underlying stochastic process , say for $dz_1$, is now of the form
\begin{equation}
\includegraphics[width=70mm]{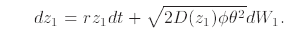} \label{26} 
\end{equation}
The solution of the one-point function $g$ is (for one asset) again a Gaussian form. The drift-free transformation $x_1 = {e^{-rt}} z_1 $ and $x_2 =  {e^{-rt}} z_2$ allows us to obtain the solution for the one-point function $g(x_i ,t), i=1$ 
\begin{equation}
\includegraphics[width=100mm]{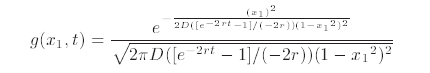} \label{27} 
\end{equation}
The choice of the boundary terms is again the simplest possible delta functions corresponding to the instantaneous measurement. Performing the integration we obtain the formal solution 

\begin{equation}
\includegraphics[width=80mm]{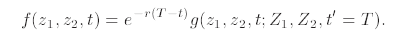} \label{28} 
\end{equation}

\section{Conclusion}
 In  this article, we have derived a risk neutral valuation model analogous to the 
 Black-Scholes and Merton derivatives valuation model. This model is derived 
 for portfolios of assets of quantum algorithms in continuous variables. The 
Fokker-Planck equations for the qbits are the starting 
 points for the derivation of the formal solutions for the derivatives which in 
 this interpretation are the distributions for the measurement of the observables
 of the quantum computation algorithms of qbits encoded on harmonic oscillators and optics. 
 risk neutral rate of evolution of the value of the portfolio $d\Pi=r\Pi dt$ and for which 
the statistics of the computation are derived for future terminal computation time $t'=T$ formal solution brought back to the current time $t$ at the known or desired rate of return. This can be interpreted in several ways, including $r$ the probability of successful computation per computation time step, The number of successful computations algorithms per unit time step, etc. and is an analogy to the rate of return on investment in finance. A connection to financial risk analysis and quantum computation is thus established, as the traditional trading strategies of risk minimization by portfolio optimization hedging schemes such as delta neutral, etc., have a direct correspondence to observation strategies.

\pagebreak 
*Email address: fnmfnm2@yahoo.com, fmicha3@uic.edu  \\

\end{document}